\documentclass[a4paper,rmp]{revtex4}
\usepackage{epsfig}
\begin{document}
\section{Simple 8085 ${\rm \mu}$p compatible I/O card}
\begin{flushright}
{\it with} Arti Dwivedi\\
\end{flushright}
\vskip 0.5cm
\begin{center}
{\bf Abstract}
\end{center}

\par {\it A simple interfacing project with the 8085-microprocessor kits 
available in under graduate college labs has been discussed. The interface card 
to study the I-V characteristics of a p-n diode emphasizes how the 
microprocessor can be used to do experiments in physics. Also, since the whole 
project was done within Rs400/- it can easily be popularized.}

\subsection{Introduction}

\par There is a Malayalam proverb, "the lazy fellow will ultimately have to 
lift the mountain". The scientist to save the trouble of doing routine 
experiments always wanted to develop a machine to do it for him, with 
the scientist free to analyze the collected data. Isn't this a symptom of a 
lazy fellow? Well these lazy fellows went on to develop the computer to 
achieve their rest. Rest assured "a lot of work". Apart from the joke, a 
computer is an important device in experiments, where the results pour in 
very very slowly or very rapidly. Consider, the difficulty of measuring the 
discharging of a capacitor in milli-seconds or in couple of hours.
The microprocessor communicates in 1's and 0's, i.e. it is only capable of 
digital communication. However, the external or outside world as we see it, 
communicates in analog. Thus, the basic requirement for the microprocessor to 
communicate with the analog world is a device capable of converting digital 
signals to analog signal and visa versa. The Input part of the device takes 
analog signals and converts it digital signals, while Output part of the device 
receives digital signals from the microprocessor and sends out analog signals. 
Together they form what is called an I/O device or more popularly an interfacing 
card. Anyone with a basic understanding of digital electronics [1, 2] would 
immediately realize that the input part would require Analog-to-Digital 
Converter (ADC) while the output part would have a Digital-to-Analog Converter 
(DAC).  In this article we describe how an ADC and DAC chip was used to develop 
a low cost 8085 microprocessor compatible interface card, which was then used to 
measure the I-V characteristics of a P-N diode. 

\subsection{Designing of the Interfacing Card}

\subsection*{Analog to Digital Converter}
\par The IC0804 ADC from National Semiconductors is a low cost chip. The cost is 
less then Rs 200/- in Delhi market. The cost is on the lower side since the 0804 
is an single channel ADC. This implies only a single source of analog signal 
can be given to the ADC. Multi-channel ADC's are available in the market, 
however, the cost and programming complexity goes up.

\begin{figure}[h]
\begin{center}
\epsfig{file=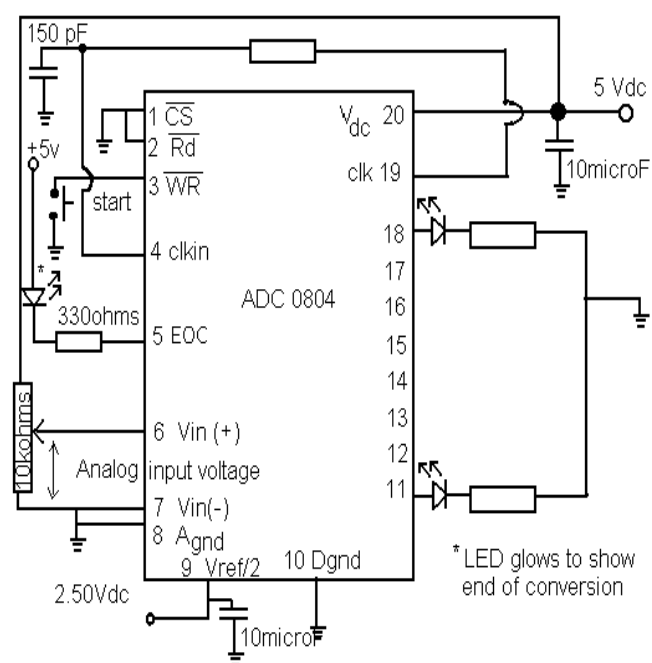,height=3.5in,width=3.5in}
\caption{Circuit using IC0804 for converting input analog voltage into 
corresponding digital signal.} 
\label{fig:arti1}
\end{center}
\end{figure}

\par The IC0804 converts analog signal to digital data by successive 
approximation method. In this method, a internal DAC keeps on comparing its 
output to the analog input. If the two voltage levels match, the DAC's input 
is the digital equivalent of the analog input. Figure 1 shows how to use/test 
an IC0804\cite{arti2}. An input analog signal varying between 0 to 5volts can 
be given at pin 6. The corresponding digital signal from ${\rm 00_H}$ to
${\rm FF_H}$ is collected from pin 11-18. Pin 11 gives the Most-Significant-Bit 
(MSB, ${\rm D_7}$) while Pin 18 gives the Least-Significant-Bit (LSB, ${\rm
D_0}$). The converter requires a clock pulse at pin 4. It is generated using a 
built in clock by connecting a resistor and capacitor externally at pin 19 and 
4. The time period of the clock pulse is given as 
\begin{eqnarray}
T=1.1RC\nonumber
\end{eqnarray}
For proper conversion the chip requires a control signal. The 
Start-Of-Conversion pulse ${\rm (\overline{SOC}}$) tells the chip to keep on 
varying the internal DAC input and compare with input analog input. When the 
chip has completed and got the answer, informs the user by giving an 
End-Of-Conversion pulse (${\rm \overline{EOC}}$). The IC in fact, at pin 5 
(${\rm \overline{EOC}}$) always gives 5volts, i.e. digital 1. When the 
conversion is finished, the signal goes low indicating completion. Care should 
be taken on selection of clock pulse (R \& C), since the end of conversion 
signals generation is sensitive to the clock pulse (the value of R and C 
selected was ${\rm 680K\Omega}$ and 150pF respectively, giving 
${\rm T=115\mu s}$). Further details of IC0804 can be downloaded from the 
National Semiconductor's web site. 

\subsection*{Digital To Analog Converter}
A DAC 0808 can be used to convert the digital input given to pins 5 to 12 of the 
chip. The analog output appears at pin 4, with the voltage level varying from 
zero to negative 5volts. An inverting amplifier is used to make the voltage 
level vary in the positive direction. The DAC0808 in itself is cheap, costing 
just over Rs 100/- in Delhi market. However, a simple resistive ladder digital 
to analog converter circuit [1] was used in this project.  The circuit, with a 
non-inverting op-amp circuit was used (whose cost was just Rs 10/-). The DAC0808 
would obviously be better in terms of accuracy. However, in a under-graduate lab, 
the resistive ladder circuit proved to be good enough.

\begin{figure}[h]
\begin{center}
\epsfig{file=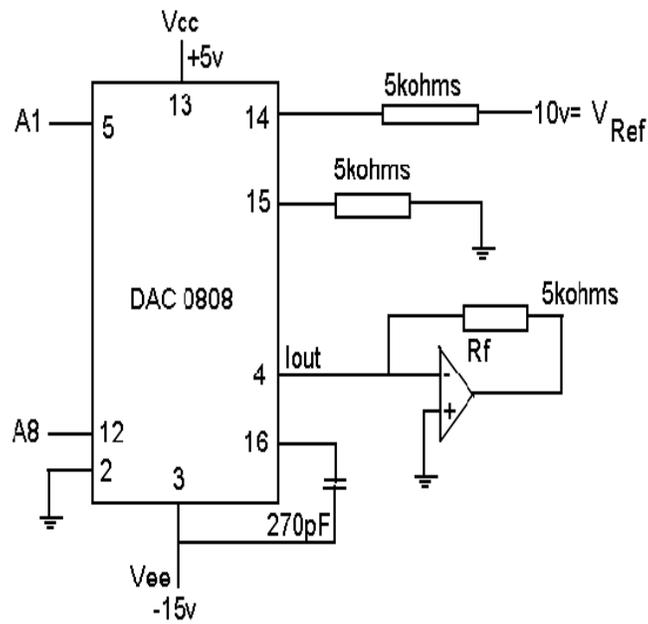,height=3.4in,width=3.4in}
\caption{A DAC0808 circuit with op-amp amplifier used to  invert the 0 to -5v 
output.} 
\label{fig:arti2}
\end{center}
\end{figure}

\subsection{The Interfacing Card}
\par After the hardware, a program has to be developed. Along with generating 
the input for the DAC and collecting the output of the ADC, the program should 
generate the ${\rm \overline{SOC}}$ signal for the ADC. The program should also 
after sending the ${\rm \overline{SOC}}$, continuously monitor for the  signal. 
All this can be achieved using either the 8155 or 8255 peripheral present on 
the 8085 microprocessor trainer kit. We selected the 8255 peripheral for the 
present project. A introduction to the 8255 peripheral chip is beyond the scope 
of this article, and details of the same can be found in Goankar\cite{arti3}. 
To appreciate the programming part of this project would require the reader to 
be familiar with 8255 peripheral, as also, the 8085 microprocessor. 

\par The automation of measuring the I-V characteristics of a diode is achieved 
by making the DAC generate a ramp signal. In simple words the output of the DAC 
would be a linearly increasing voltage with time, the voltage level going from 
0v to 5v. This is done by the microprocessor counting from ${\rm 00_H}$ to
${\rm FF_H}$. With each increasing count, the DAC's output voltage increases by 
19.6mV (5/255volts). This voltage is given as an input signal to the circuit 
shown below. 

\begin{figure}[h]
\begin{center}
\epsfig{file=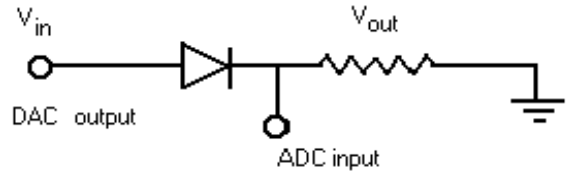,height=1in,width=3in}
\caption{Circuit for measuring IV characteristics using interface card.} 
\label{fig:arti3}
\end{center}
\end{figure}

\par The voltage across the ${\rm 1K\Omega}$ resistance is converted to digital 
values and is collected and stored by the microprocessor. So the program 
increments the count and sends the digital data to the DAC which is used as the 
input voltage for the diode circuit. The voltage across the resistance is 
converted to digital signal and stored in a memory location of the 
microprocessor. This goes on in a loop to completely obtain the diodes I-V 
characteristics. Table I lists the required program to achieve the above 
objective.  

\par Before relying on the results of the device, it is necessary to see the 
linearity of the device. This is done by giving the output of the DAC as 
input of the ADC. The program remains the same. Figure 4 shows the device 
to be appreciably linear. The listed program above saves the input to the 
DAC and the output of the ADC. As seen from figure 3, the voltage output of 
the ADC is a measure of the circuit's current. To plot the I-V characteristics, 
the ADC's results are plotted on the 'y' axis (${\rm V_r}$, the voltage drop 
across the ${\rm 1K\Omega}$ resistor, which is proportional to the current) 
while this data has to be subtracted from the DAC's output for voltage across 
the diode ('x'-axis). Figure 5 shows the I-V characteristics of a 1N407 diode 
measured with the designed interface card. The knee voltage is evident, just 
more then 0.4v. 

\begin{figure}[h]
\begin{center}
\epsfig{file=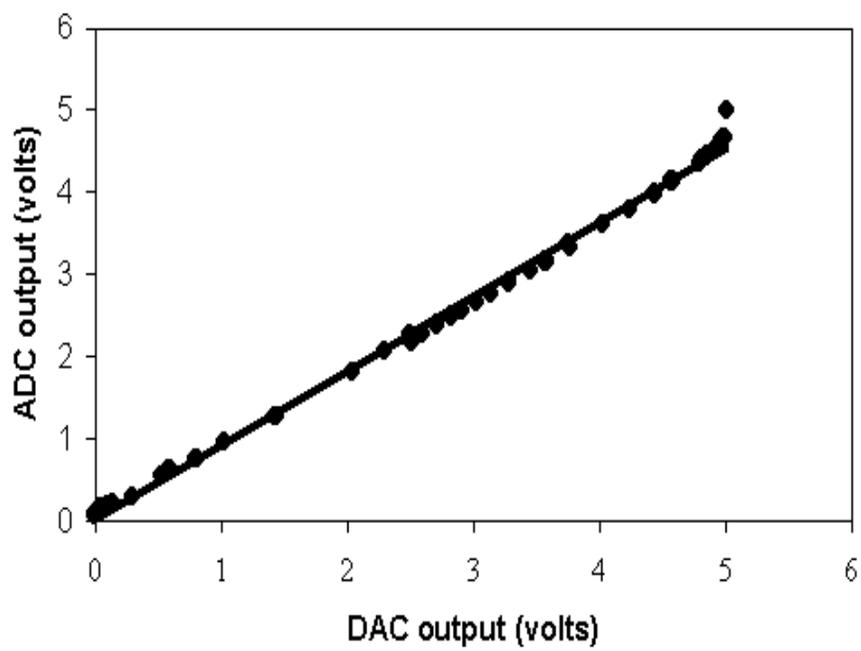,height=3.4in,width=4.5in}
\caption{The interface card was found to be perfectly linear, i.e. the output 
of the ADC is equal to the input given to the DAC.} 
\label{fig:arti4}
\end{center}
\end{figure}

\par Obviously, the prospect of copying 512 (${\rm =2\times 256}$) data from 
the microprocessors memory might put of the reader. However, with a small 
additional program and by connecting the DAC to an oscilloscope, the data can 
be read and displayed on the CRT. 

\par This project was done in a graduate college laboratory, with a small budget of 
Rs 400/-. A reader who has knowledge of computers and of computer programming 
(especially of C programming) would benefit from Probhir Goyal's 
article\cite{arti4}, if he or she were interested for developing an interfacing 
card for the computer.

\begin{figure}[h]
\begin{center}
\epsfig{file=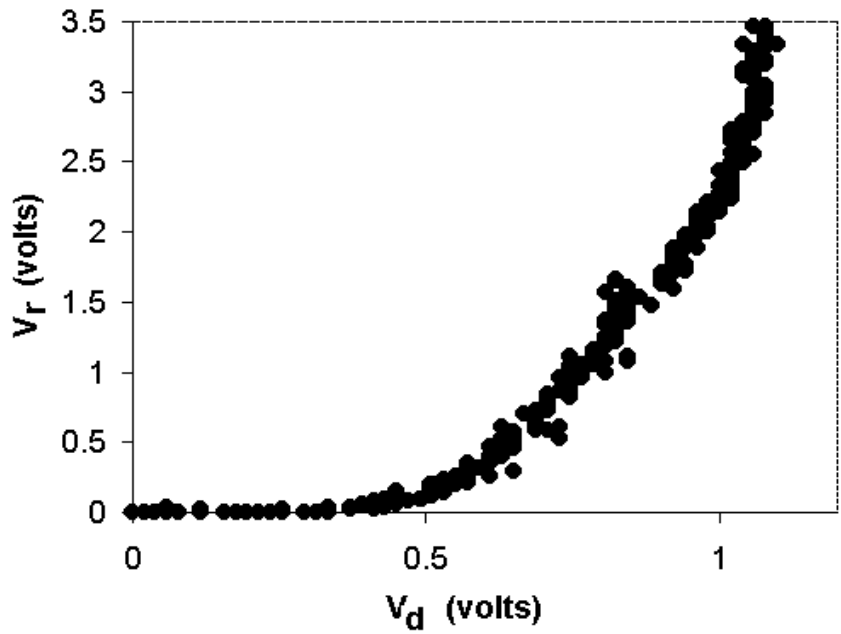,height=3.4in,width=4.5in}
\caption{The I-V characteristics of a p-n diode as measured by the designed 
interface card.} 
\label{fig:arti5}
\end{center}
\end{figure}

\pagebreak
\begin{table}
Program required for the interface card.
\begin{center}
\begin{tabular}{cccc}\hline\hline\\
Memory address & Hex Code & Instruction & Comments\\
\hline\hline\\
C000 &	26 &	MVI H		& Initialize counter H to 0\\
C001 &	00 &	${\rm 00_H}$	& \\
C002 &	06 &	MVI B		& Initialize counter B to 0\\
C003 &	00 &	${\rm 00_H}$	& \\
C004 &	3E &	MVI A		& "Load control word in control \\
C005 &	98 &	${\rm 98_H}$	& Register"\\
C006 &	D3 &	OUT		& \\
C007 &	13 &	${\rm 13_H}$	& \\
C008 &	78 &	MOV A, B	& \\
C009 &	D3 &	OUT		& \\
C00A &	11 &	${\rm 11_H}$	& \\
C00B &	11 &	LXI D		& "Setup DE pair as pointer \\
C00C &	00 &	00		& for destination memory C100"\\
C00D &	C1 &	C1		& \\
C00E &	7B &	MOV A, E	& Move data of E into A\\
C00F &	84 &	ADD H		& Add data in H to A\\
C010 &	5F &	MOV E, A	& Move data of A into E\\
C011 &	78 &	MOV A, B	& Move into A data of B\\
C012 &	12 &	STAX D		& Store data in pointed memory\\
C013 &	3E &	MVI A		& "Load BSR mode control\\
C014 &	00 &	${\rm 00_H}$	& Word in control register \\
C015 &	D3 &	OUT		& To reset PC0"\\
C016 &	13 &	${\rm 13_H}$	& \\
C017 &	0E &	MVI C		& Wait in delay loop\\
C018 &	15 &	${\rm 15_H}$	& \\
C019 &	0D &	DCR C		& \\
C01A &	C2 &	JNZ		& \\
C01B &	19 &	19		& \\
C01C &	C0 &	C0		& \\
C01D &	3E &	MVI A		& "Load BSR mode control\\
C01E &	01 &	${\rm 01_H}$	& word in control register to\\
C01F &	D3 &	OUT 		& PC0"\\
C020 &	13 &	${\rm 13_H}$	& \\
C021 &	DB &	IN		& Read port C\\
C022 &	12 &	${\rm 12_H}$	& \\
C023 &	17 &	RAL		& Move PC7 into carry flag\\
C024 &	DA &	JC		& Wait in loop till EOC is low\\
C025 &	21 &	21		& \\
C026 &	C0 &	CO		& \\
C027 &	DB &	IN		& "Read Port A (ADC output)\\
C028 &	10 &	${\rm 10_H}$	& Into accumulator"\\
C029 &	4F &	MOV C, A	& Move data from A into C\\
C02A &	11 &	LXI D		& "Setup DE pair as a pointer\\
C02B &	00 &	00		& For destination memory"\\
C02C &	C2 &	C2		& \\
C02D &	7B &	MOV A, E	& Move data in E into A\\
C02E &	84 &	ADD H		& Add the data in H into A\\
C02F &	5F &	MOV E, A	& Move data in A into E\\
C030 &	79 &	MOV A, C	& Move data in C into A\\
\hline\hline\\
\end{tabular} 
\end{center}
\end{table}

\begin{table}
\begin{center}
\begin{tabular}{cccc}\hline\hline\\
Memory address & Hex Code & Instruction & Comments\\
\hline\hline\\
C031 &	12 &	STAX D		& Store ADC o/p into memory\\
C032 &	04 &	INR B		& Increment register B\\
C033 &	24 &	INR H		& Increment register H\\
C034 &	0E &	MVI C		& Wait in delay loop\\
C035 &	FF &	${\rm FF_H}$	& \\
C036 &	OD &	DCR C		& \\
C037 &	C2 &	JNZ		& \\
C038 &	36 &	36		& \\
C039 &	C0 &	C0		& \\
C03A &	78 &	MOV A, B	& Move data in B into A\\
C03B &	FE &	CPI		& Compare data in A with ${\rm 00_H}$\\
C03C &	00 &	${\rm 00_H}$	& \\
C03D &	C2 &	JNZ		& "If data in B is not 0 go back  \\
C03E &	04 &	04		& In loop"\\
C03F &	C0 &	C0		& \\
C040 &	76 &	HLT		& End Of  Program\\
\hline\hline\\
\end{tabular} 
\end{center}
\end{table}

\pagebreak

\end{document}